\documentclass[11pt,twoside]{article}
\usepackage{asp2010,bm,url}

\newcommand{\et}{\,et\,al.\,}

\resetcounters

\begin{document}

\title{Excitation of stellar pulsations}
\author{G\"unter Houdek
\affil{Institute of Astronomy, University of Vienna, 1180 Vienna, Austria}}

\begin{abstract}
In this review I present an overview of our current understanding of 
the physical mechanisms that are responsible for the excitation of 
pulsations in stars with surface convection zones. 
These are typically cooler stars such as the 
Delta Scuti stars, and stars supporting solar-like oscillations.
\end{abstract}

\section{Introduction}
Convection dynamics affects the driving and damping of pulsations particularly
in stars with convectively unstable outer layers, including also the F stars 
which have rather shallow surface convection zones and in which solar-like 
oscillations have been observed. The type of stars in which the coupling
between pulsation and the turbulent velocity field is important are, for
example, the rapidly oscillating Ap stars, Delta Scuti stars, 
$\gamma$ Doradus, RR Lyrae stars, and stars which support stochastically excited
oscillations. In this review I shall only address two of them, Delta
Scuti stars and solar-type stars. 

\section{Delta Scuti stars}
Delta Scuti ($\delta$ Sct) stars have masses $M$ between 
1.5$\,\la\,M\,/\,{\rm M}_\odot\!\la\,$2.5 and 
luminosities $L$ between 0.6$\,\la\log L\,/\,$L$_\odot\la\,$2.0. They are in the central or 
shell hydrogen burning phase with complex multiperiodic oscillation spectra, including 
both radial and nonradial modes (e.g., 79 pulsation modes have been detected in FG Vir,
Breger \et 2005). The oscillations are low-order p modes with relatively small amplitudes
ranging from 10$^{-3}$ -- 0.1\,mmag and with periods between 18\,min and 8\,h. The modes
are driven by the kappa mechanism in the second stage of helium ionization.
Cooler $\delta$ Sct stars have substantial surface convection zones and, similarly
to solar-like stars, the stability properties of the oscillation modes are crucially
affected by the convection dynamics. In particular, the return to stability
of low-order p modes at the cool boundary of the classical instability strip is
predicted only with the inclusion of convection dynamics in the stability analysis
(e.g., Houdek \et 1999, Houdek 2000, Xiong \& Deng 2001, Dupret \et 2005).
These authors use different implementations for modelling the interaction of the 
turbulent velocity field with the pulsation. The currently most commonly used
time-dependent convection formulations for studying the stability of stellar pulsations
are those by Gough (1965, 1977a,b), Xiong (1977, 1989) and Grigahc{\`e}ne \et (2005) the
latter being a generalization of Unno's (1967) formulation. Although they all
adopt the Boussinesq approximation to the fluid equations they differ substantially 
in detail. A brief
discussion of the main differences between these convection models was recently 
given by Houdek (2008). All three convection formulations model successfully
the location of the red edge of the classical instability strip (IS), however, the
physical mechanism responsible for the return to stability is very different
in all three computations. Dupret \et (2005) report that it is predominantly
the convective heat flux, Xiong \& Deng (2001) the viscosity of the
small-scale turbulence, and Houdek (2000) that it is predominantly the
momentum flux (turbulent pressure $p_{\rm t}$) that stabilizes the pulsation 
modes at the red edge of the IS. 
The results of these stability calculations
are illustrated in Figure~1 in terms of accumulated work integrals $W$ 
for stellar models located just outside the cool edge of the IS, i.e.
for models for which the pulsation is found to be stable. 


\begin{figure}[!ht]
\label{f:1}
\plotone{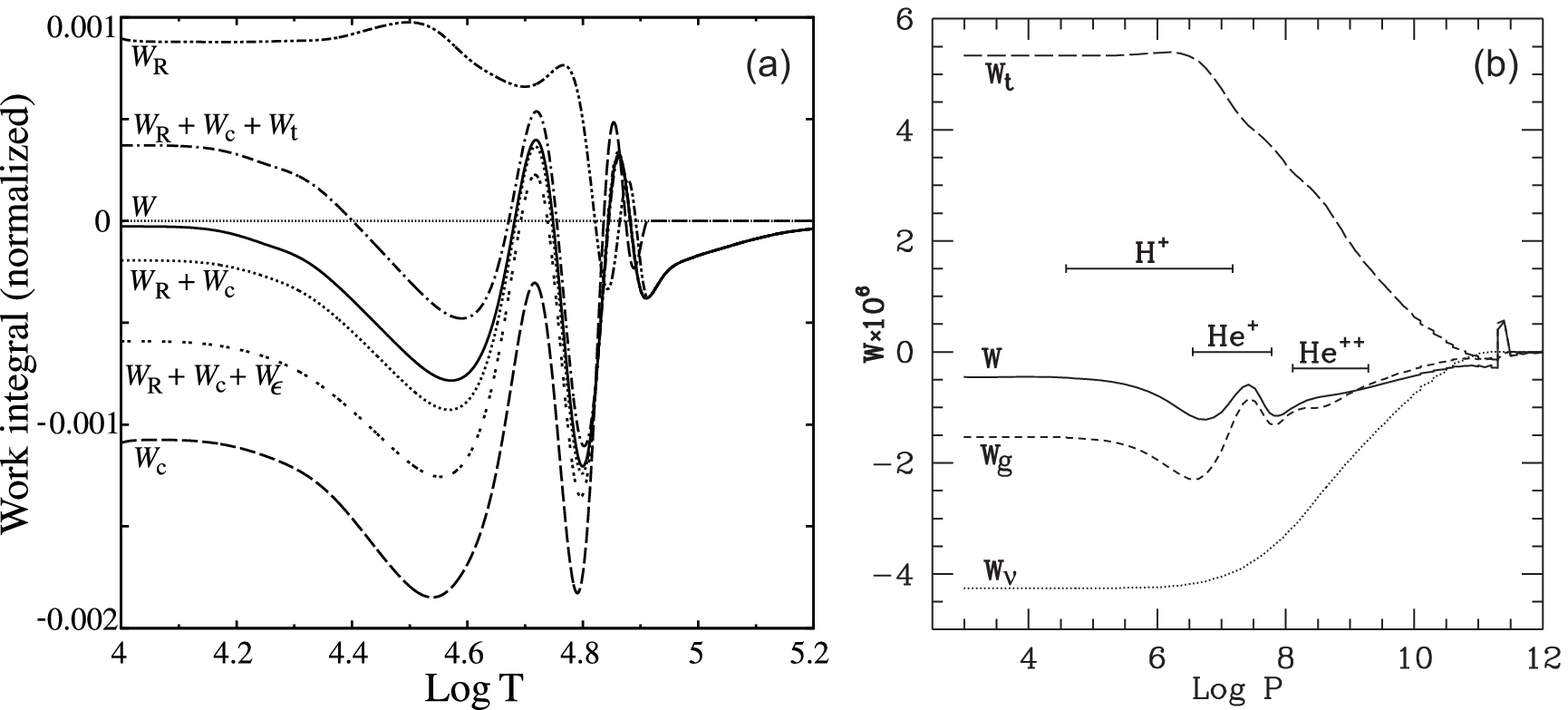}
\plotfiddle{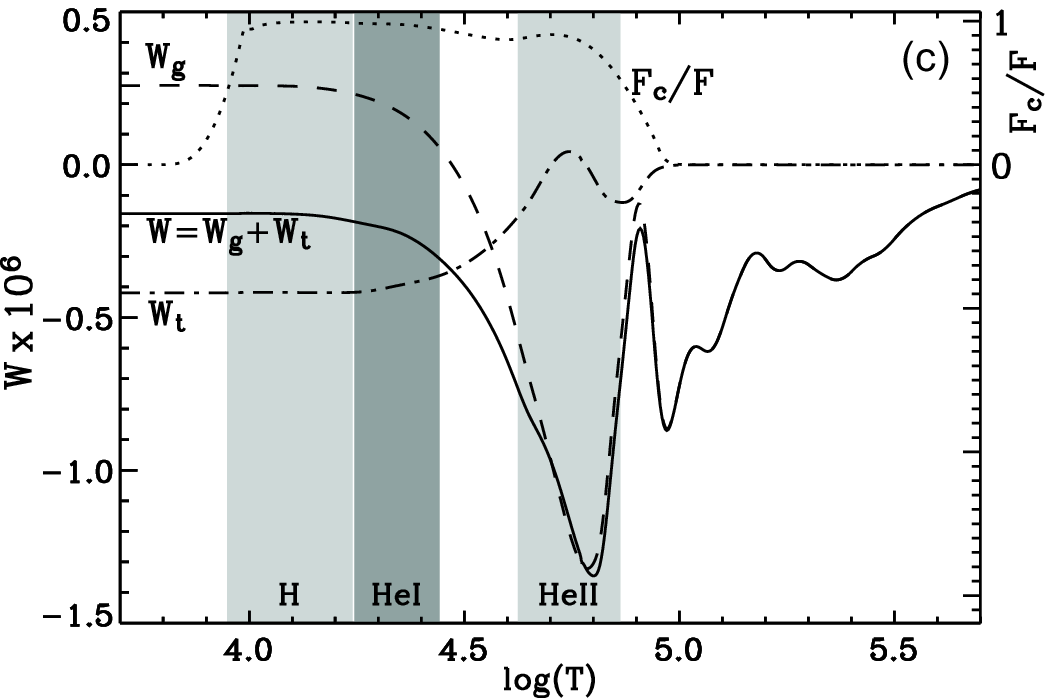}{7.0truecm}{0}{65}{65}{-173pt}{75pt}
\parbox{\textwidth}{\hspace{7.0cm}
   \parbox{6.6cm}{
      \vspace{-103truemm}
      \caption{\sloppy\hspace{-5mm}
       Accumulated work integrals of $\delta$ Sct models located 
       just outside the
       red edge of the classical instability strip, i.e. the
       total work integral $W<$0. Panel (a) shows
       the results by Dupret\et (2005) using the time-dependent
       mixing-length model based on Unno's (1967) and Gabriel's (1987, 1996) 
       descriptions, and extended by Grigahc{\`e}ne\et (2004). 
       Panel (b) illustrates the results reported by
       Xiong \& Deng (2007) using Xiong's
      }
   }
}
\parbox{\textwidth}{\vspace{-130pt}\hspace{20pt}
       \parbox{12.1cm}{\small
       (1989) Reynolds-stress-model-like time-dependent convection
       formulation.
       Panel (c) are the results by Houdek (2000) using Gough's (1977a,b)
       nonlocal, time-dependent mixing-length model. Panel (a) is adapted
       from Dupret\et (2005), panel (b) from Xiong \& Deng (2007), and (c) from
       Houdek (2000).}
}
\vspace{-60pt}
\end{figure}


Nonadiabatic pulsation computations provide complex eigenfunctions and eigenfrequencies 
$\omega=\omega_{\rm r}+{\rm i}\omega_{\rm i}$ for modes with 
angular frequencies $\omega_{\rm r}$ and growth/damping rates $\omega_{\rm i}$.
Work integrals $W$ for complex eigenfrequencies were, for example, 
provided by Baker \& Gough (1979) and can be divided into contributions
arising from terms associated with thermal energy equation 
(e.g., gas pressure perturbation, $W_{\rm g}$), and terms associated 
with the momentum equation (e.g., momentum flux perturbation, $W_{\rm t}$).

 
 \begin{figure}[!ht]
 \plotone{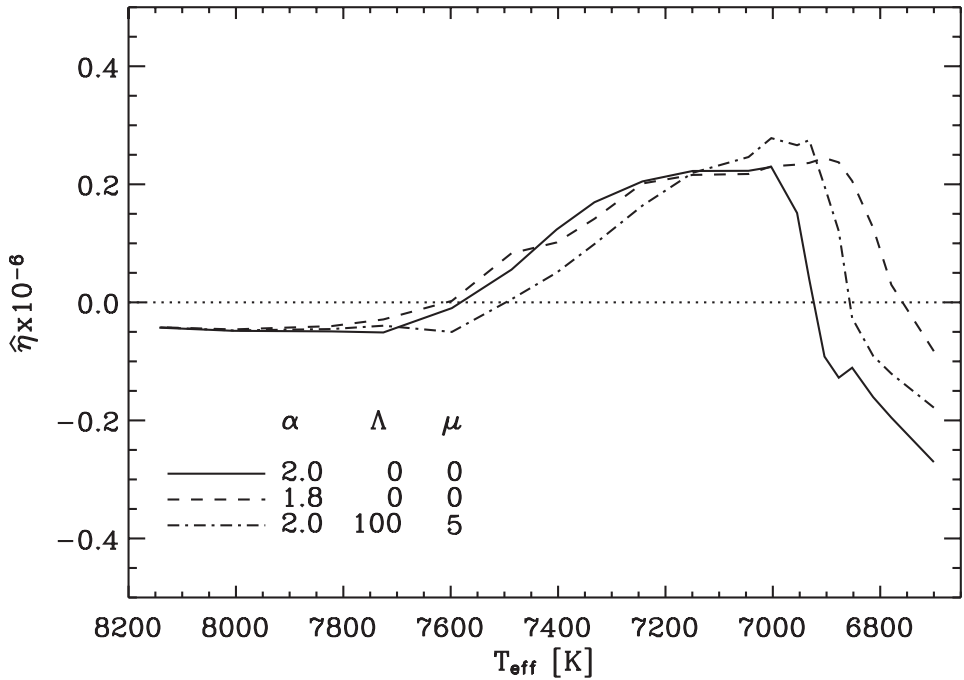}
 \caption{Linear stability coefficients $\widehat\eta:=\omega_{\rm i}/\omega_{\rm r}$
          (positive values indicate instability) of 
          the fundamental radial acoustic mode for an 1.7 M$_\odot$ $\delta$ Sct 
          star in the hydrogen-burning phase crossing the instability strip. 
          Results are shown for two different mixing-length parameters and for
          a calculation in which the effect of acoustic radiation in the equilibrium
          model was included according to expression~(\ref{e:acoustic-radiation}).
          }
 \label{f:2}
 \end{figure}
 

The upper left panel (a) of Figure~1 shows the results of 
Dupret \et (2005). The individual contributions to the total $W$ arising 
from the radiative flux, $W_{\rm R}$, the convective flux, $W_{\rm c}$, the turbulent
pressure, $W_{\rm t}$ ($W_{\rm R}+W_{\rm c}+W_{\rm t})$, and from the
perturbation of the turbulent kinetic energy dissipation, $W_\epsilon$
($W_{\rm R}+W_{\rm c}+W_{\epsilon}$) are indicated by different line styles. 
It demonstrates the near cancellation effect
between the contributions of the turbulent kinetic energy dissipation, $W_{\epsilon}$,
and turbulent pressure, $W_{\rm t}$, making the contribution from the fluctuating
convective heat flux, $W_{\rm c}$, the dominating damping term.

The results of the calculations by Xiong \& Deng (2001) are displayed in 
the upper right panel (b) of Figure~1. Contributions to the total 
$W=W_{\rm g}+W_{\rm t}+W_{\nu}$ arising from the gas pressure 
fluctuations, $W_{\rm g}$, the turbulent pressure
fluctuations, $W_{\rm t}$, and from an effective
viscosity from the small-scale turbulence, $W_{\nu}$, are 
indicated by different
line styles. The ionization zones of hydrogen (H) and helium (He)
are indicated. In this calculation the dominating agent for making
the pulsations stable is the damping contribution from the small-scale
turbulence, $W_{\nu}$. 

Panel (c) of Figure~1 shows the results of Houdek (2000) using Gough's (1977a,b)
convection formulation. Contributions to $W=W_{\rm g}+W_{\rm t}$ (solid curve) 
arising from the gas pressure perturbation, $W_{\rm g}$ (dashed curve), and the 
turbulent pressure fluctuations, $W_{\rm t}$ (dot-dashed curve), are indicated. 
The dotted curve is the ratio of the convective to the total heat flux, $F_{\rm c}/F$, 
and the ionization zones of H and He (5\% to 95\%) are indicated. In this
calculation it is the contribution from the turbulent pressure fluctuation, $W_{\rm t}$,
that stabilizes the pulsation mode.

It is interesting to note that all three convection descriptions include, although in
different ways, the perturbations of the turbulent fluxes, but Gough (1977a), 
Xiong (1977, 1989) and Unno\et (1989) did not include the contribution $W_\epsilon$ to 
the work integral because in the Boussinesq approximation (Spiegel \& Veronis 1960) the
viscous dissipation is neglected in the thermal energy equation. In practise, however, this
term may be important. 

All three convection models need calibration for the mixing-length parameter $\alpha$.
The effect of varying $\alpha$ on the location of the IS
for the fundamental radial acoustic mode is illustrated in Figure~\ref{f:2} for
an 1.7$\,$M$_\odot$ star in the central hydrogen-burning phase. Decreasing $\alpha$ 
results in a significant shift of the cool boundary of the IS
towards lower surface temperatures, whereas the blue edge is essentially unaffected. This
comes about because reducing $\alpha$ diminishes the stabilizing effect of the
turbulence on the pulsations, which is, however, offset by the increasing efficacy
of the surface convection with decreasing surface temperature $T_{\rm eff}$ of the star.
The observed location of the cool edge of the IS can therefore by used to calibrate
$\alpha$ in $\delta$ Sct stars.

Another effect, yet neglected in essentially all stability computations, is the effect
of acoustic radiation in the equilibrium model. The generation of acoustic waves
is omitted in convection models which assume either the anelastic (Gough 1969)
or Boussinesq approximation (Spiegel \& Veronis 1960) to the fluid equations.
However, through the generation of sound waves, kinetic energy from the turbulent 
motion will be converted into acoustic radiation (Lighthill 1952) and thus reduce 
the efficacy with which the motion might otherwise have released potential energy
originating from the buoyancy forces. This effect may become important in stars
with large turbulent Mach numbers $M_{\rm t}:=w/c$, where $w$ is the vertical rms 
component of the convective velocity field $\bm{u}=(u,v,w)$, and
$c$ is sound speed. In the phenomenological picture of an overturning convective
fluid element (eddy), which maintains balance between buoyancy forces and
turbulent drag by continuous exchange of momentum with other elements
and its surrounding (e.g., Unno 1967), the equation of motion for
the turbulent element of vertical size $\ell$ can be written as
\begin{equation}
\frac{2w^2}{\ell^2}=g\widehat\alpha T^\prime - \frac{P_{\rm ac}}{\rho w}\, ,
\label{e:acoustic-radiation}
\end{equation}
where $T^\prime$ is the Eulerian temperature fluctuation, 
$\widehat\alpha$ the coefficient of thermal expansion, and
$g$ is the acceleration due to gravity. 
The rate of energy of acoustic radiation per unit volume, $P_{\rm ac}$ is
estimated according to the Lighthill-Proudman formula
\begin{equation} 
P_{\rm ac}=\Lambda\frac{\rho w^3}{\ell}\left(\frac{w}{c}\right)^\mu\, ,
\end{equation} 
where $\Lambda$ and $\mu$ are the coefficients for emissivity and Mach-number 
dependence. If the emission of acoustic waves by homogeneous, isotropic turbulence
is dominated by the largest convective eddies, the acoustic emission $P_{\rm ac}$
scales with a Mach-number dependence $\mu=5$ (Lighthill 1952) and the
emissivity coefficient $\Lambda\simeq100$ for a solar model (Stein 1968), which
we also adopt here in our estimate for $P_{\rm ac}$.

The effect of the acoustic radiation $P_{\rm ac}$ in the mean model on the location 
of the instability strip for the fundamental radial acoustic mode is illustrated by 
the dot-dashed curve in Figure~\ref{f:2} for an 1.7$\,$M$_\odot$ star in the central 
hydrogen-burning phase. At the cool edge the effect of including $P_{\rm ac}$ is 
similar to reducing the mixing length, however, at the hot (blue) edge of the IS 
$P_{\rm ac}$ has a much larger effect than changing the vertical extent of the
convective elements.
 
\section{Solar-like stars}
Kepler has been providing data of solar-like oscillations with unprecedented 
quality in more than 2000 stars  
(e.g., Chaplin\et 2011; Garc{\`i}a\et 2011; Huber\et 2011b). One of
the many important outcomes of the Kepler data analyses is the confirmation of
the problem of reproducing theoretically the 
observed amplitudes of stochastically excited modes in stars hotter than the Sun. 
In general, Kepler observed photometric amplitudes in stars located near the red 
edge of the IS that are rather similar to those found in the Sun, whereas
theoretical amplitude estimates are up to 3-4 times larger than the Kepler values.
This discrepancy in the oscillation amplitudes was already recognised before
(Houdek 2006) when first
ground-based spectroscopic observations were available, such
as for the F-star Procyon (e.g. Martic\et 1999, Arentoft\et 2008).
It therefore behoves us to address the possible reasons why our current theory 
fails to reproduce observed amplitudes of solar-like oscillations in hotter 
stars. Reviews on the theory and modelling of stochastically excited modes
were recently given, for example, by Appourchaux\et (2009) and Houdek (2010).
Therefore I shall summarize only the most important matters for our
discussion.

Solar-like oscillations are intrinsically damped but driven stochastically
by the vigorous turbulent convection in the very outer stellar layers. The
height $H$ of a single peak in the observed oscillation power spectrum 
can be obtained from taking the Fourier transformation of a damped,
harmonic oscillator for the surface displacement followed by an 
integration over frequency to obtain the total mean energy $E$ in a
single pulsation mode with (normalized) inertia $I$ (e.g. Chaplin\et 2005; Houdek 2006).
Provided that the observing time is long compared to the mode lifetime
the height $H$ in units of cm$^2\,$s$^{-2}\,$Hz$^{-1}$ is given by
\begin{equation}
H=\frac{2E}{\eta I}=\frac{P}{\eta^2I}\,,
\label{e:height}
\end{equation}    
where $\eta$ is the damping rate or inverse of the mode lifetime, and
$P$ is the energy-supply rate in erg$\,$s$^{-1}$. 
If only the Reynolds stress driving term is considered in the equations of
motions, the energy-supply rate $P$ for radial models is given by 
(e.g. Chaplin\et 2005)
\begin{equation}
P=
    \frac{\pi}{9I}
    \int_0^R \ell^3
    \left(\Phi\Psi rp_{\rm t}\frac{\partial{\xi}_{r}}{\partial r}\right)^2
    {\cal S}(\omega_{\rm r}; r)\,{\rm d}r\,,
\label{e:energy-supply-rate}
\end{equation}
with
\begin{equation}
{\cal S}(\omega_{\rm r}; r)=\int_0^\infty \kappa^{-2}\tilde E^2(\kappa)
                     \tilde\Omega(\tau_k;\omega_{\rm r})\,{\rm d}\kappa\,,
\label{e:frequency-factor}
\end{equation}
where $R$ is surface radius, 
$p_{\rm t}=\langle\rho ww\rangle$ is the $(r,r)$ component of the 
(mean) Reynolds stress tensor (also known as turbulent pressure; 
angular brackets denote an ensemble average), 
$\xi_r$ is the normalized radial part of the pulsation eigenfunction $\bm{\xi}$,
and the product of $\Phi$ and $\Psi$ is a factor 
of unity accounting for the anisotropy of the turbulent velocity field $\bm{u}$.
The spectral function ${\cal S}$ accounts for contributions to $P$
from the small-scale turbulence and includes the normalized spatial
turbulent energy spectrum $\tilde E(k)$ and the
frequency-dependent factor $\tilde\Omega(\tau_k;\omega_{\rm r})$;
$\tau_{k}:=\lambda/ku_k$ is the correlation time 
scale of eddies with wave number $k$ and velocity $u_k$ ($\lambda$ is a 
factor of order unity and accounts for uncertainties
in defining $\tau_k$), and $\kappa=k\ell/\pi$. For $\tilde E(k)$
it has been common to adopt, for example, the Kolmogorov (Kolmogorov 1941)
spectrum. For the frequency-dependent factor
$\Omega(\tau_k;\omega)$, however, which is used for evaluating the self-convolution 
$\tilde\Omega(\omega_{\rm r},\tau_k;r)=
\int\Omega(\omega,\tau_k;r)\Omega(\omega_{\rm r}\!-\!\omega,\tau_k;r)\,{\rm d}\omega$
in equation\,(\ref{e:frequency-factor}), no satisfactory theory exists. 
The two commonly adopted forms are\\
-- the Gaussian factor (Stein 1967)\,,
\begin{equation}
\Omega_{\rm G}(\omega;\tau_k)=
    \frac{\tau_k}{\sqrt{2\pi}}\,{\rm e}^{-(\omega\tau_k/\sqrt2)^2}\,;
\label{e:gaussian}
\end{equation}
-- the Lorentzian factor
   (Gough 1977b; Samadi et al. 2003; Chaplin et al. 2005)\,,
\begin{equation}
\Omega_{\rm L}(\omega;\tau_k)=
 \frac{\tau_k}{\pi\sqrt{2\ln2}}\,\frac{1}{1+(\omega\tau_k/\sqrt{2\ln2})^2}\,.
\label{e:lorentzian}
\end{equation}

\begin{figure}[!ht]
\plotone{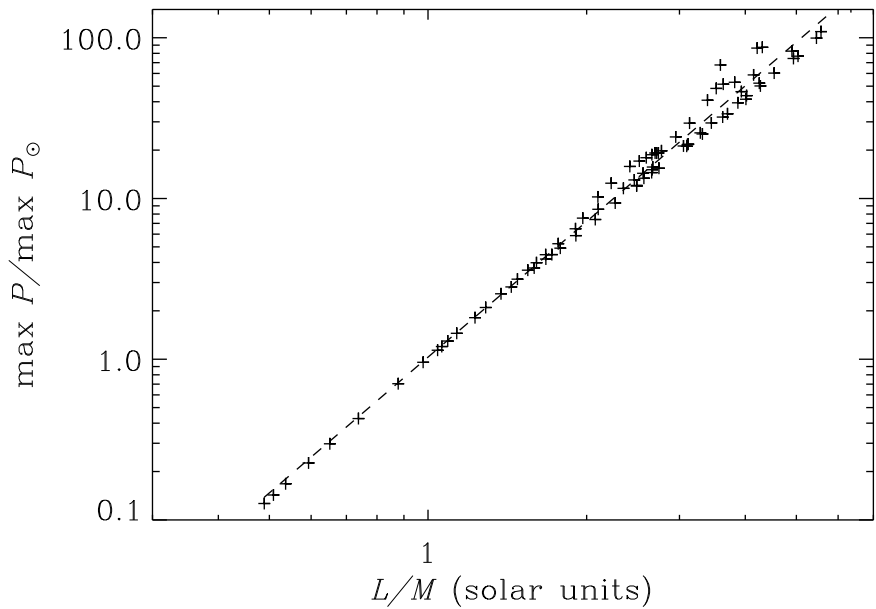}
\caption{Maximum energy-supply rates $P$ (equation~\ref{e:energy-supply-rate}) of stochastically 
excited oscillations in 83 (symbols) solar-like stellar models, computed in the manner
of Chaplin\et (2005) using a Gaussian frequency factor 
(equation~\ref{e:gaussian}). The dashed line is a fit to the 83 model data by
linear least squares, leading to an exponent $s=2.80$ in the scaling relation
max$(P)/{\rm max}(P_\odot)\sim(L/M)^s$, where $L$ and $M$ are in solar units 
(see also Table~1).
}
\label{f:3}
\end{figure}


\begin{table}[!ht]
\caption{Comparison of different stochastic excitation calculations. The correlation 
parameter $\lambda$ accounts for uncertainties in defining the correlation time 
scale of the turbulent eddies. The value of $s$ is obtained from 
fitting $(L/M)^s$ (in solar units) to the computed values of 
max$(P)/$max$(P_\odot)$ of 83 solar-like stellar models by linear 
least squares (see Figure~\ref{f:3}).}
\smallskip
\begin{center}
{\small
\begin{tabular}{llll}
\tableline
\noalign{\smallskip}
\ \ \ Excitation model & \qquad\qquad frequency factor $\Omega$& $\ \lambda$ &\ \ \ $s$ \\
\noalign{\smallskip}
\tableline
\noalign{\smallskip}
Chaplin\et (2005) & Gaussian (Stein 1967)              & 1.0 & 2.80 \\
Samadi\et (2003)  & mod. Lorentzian (Belkacem\et 2010) & 1.5 & 2.87 \\
Samadi\et (2003)  & mod. Lorentzian (Belkacem\et 2010) & 1.0 & 2.72 \\
\noalign{\smallskip}
\tableline
\end{tabular}
}
\end{center}
\end{table}

The Lorentzian frequency factor is a result predicted for the largest,
most-energetic eddies by the time-dependent mixing-length formulation
of Gough~(1977b), which decays more slowly
with depth $z$ and frequency $\omega$ than the Gaussian factor.  
Consequently a substantial fraction to the integrand of
equation~(\ref{e:energy-supply-rate}) arises from eddies situated in the deeper
layers of the Sun, resulting in a larger acoustic excitation
rate $P$. Samadi\et (2003)  reported that
Stein \& Nordlund's hydrodynamical simulations also suggest a Lorentzian
frequency factor. 
However, Chaplin\et (2005) reported that
the Lorentzian time-correlation function leads to overestimated heights $H$
at low frequencies for solar p modes. 
Recently, Belkacem\et (2010) suggested to use a modified
Lorentzian frequency factor, by cutting off 
the high-frequency tail, i.e. setting it to zero, for frequencies
$\omega>\omega_{\rm E}$. The cutoff frequency $\omega_{\rm E}$ is associated with
the inverse of the Eulerian microscale (Tennekes \& Lumley 1972) and can be
estimated with the help of Kaneda's (1993) 'random sweeping model 
approximation'.

One of the many ongoing projects of the KASC (Kepler Asteroseismic Science Consortium) 
working group 1 ``Solar-like p-mode oscillators'' 
is the comparison of various stochastic excitation models. 
In a preliminary exercise a grid of 83 solar-type main-sequence models 
was computed in the manner of Chaplin\et (2005) from which 
energy-supply rates $P$ were computed with the excitation models by 
Samadi\et (2001, 2003), using a Lorentzian frequency factor modified 
in the manner of Belkacem et al. (2010), and the excitation model by Chaplin\et (2005)
using a Gaussian frequency factor. All three computations assumed the same
equilibrium models, pulsation eigenfunctions and mode inertia. The maximum values of
the energy-supply rate $P$ for the 83 solar-like stars with masses between
0.9$\,\le\,M\,/\,{\rm M}_\odot\!\le\,$1.5, evaluated with the
excitation model of Chaplin\et (2005, see equation~\ref{e:energy-supply-rate}), 
are plotted in Figure~{\ref{f:3}} as a function of $L/M$,
where $L$ and $M$ are in solar units.
The maximum values of $P$ scale with $(L/M)^s$, where $s=2.80$.
The values for the exponent $s$ for three different excitation model calculations, 
all using the same equilibrium model and pulsation eigenfunctions $\bm{\xi}$, 
are listed in Table~1. It interesting to note that the value of $s$
varies only marginally between the different excitation models considered here, 
suggesting that the main problem of modelling the observed oscillation heights $H$
(see expressions \ref{e:height}--\ref{e:frequency-factor})
in hotter main-sequence stars is very likely related to a failure
in modelling either the equilibrium structure (e.g. the convective velocity field
and consequently $p_{\rm t}$), the pulsation 
eigenfunctions $\bm{\xi}$, or the mode damping rates $\eta=-\omega_{\rm i}$.


\begin{figure}[!ht]
\plottwo{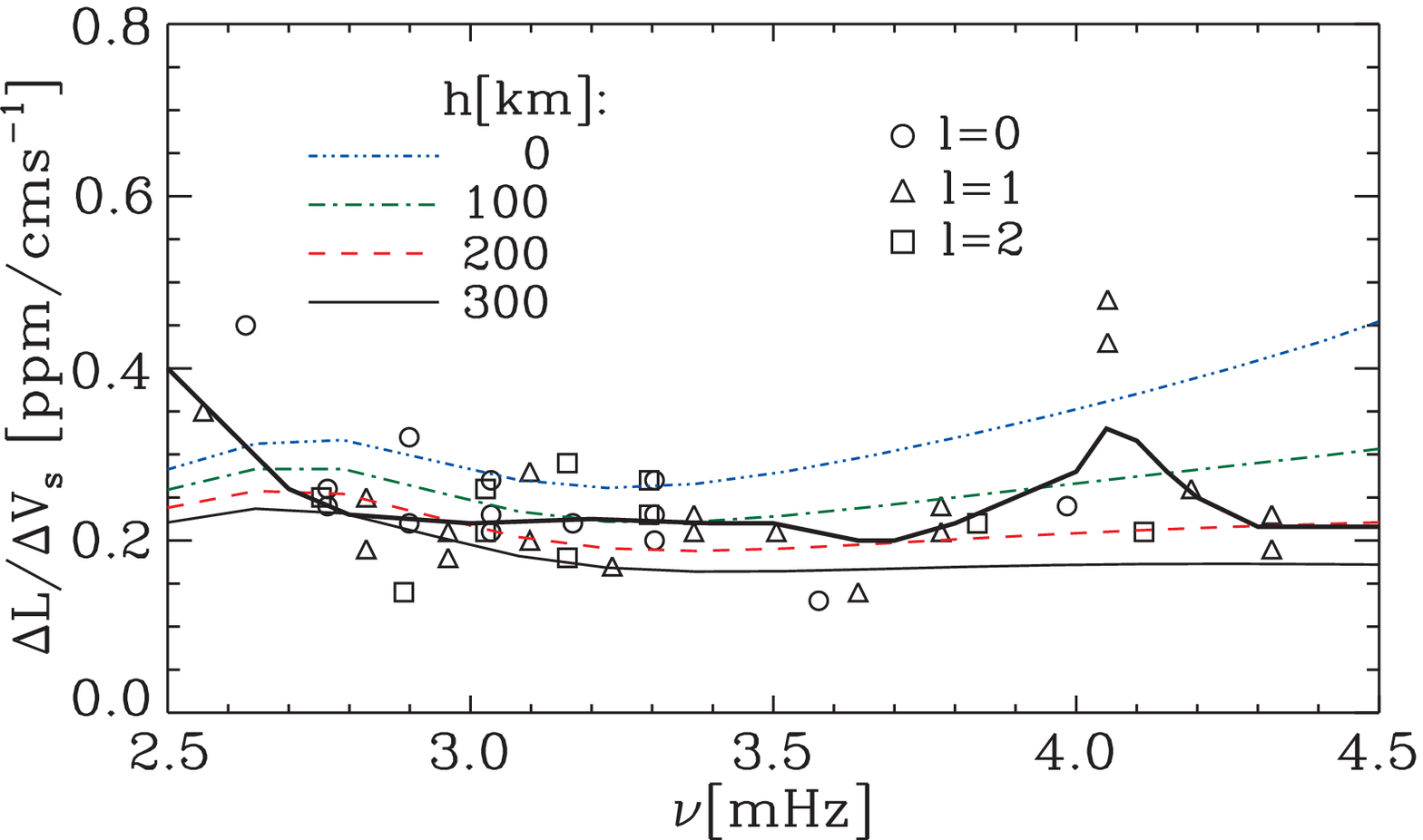}{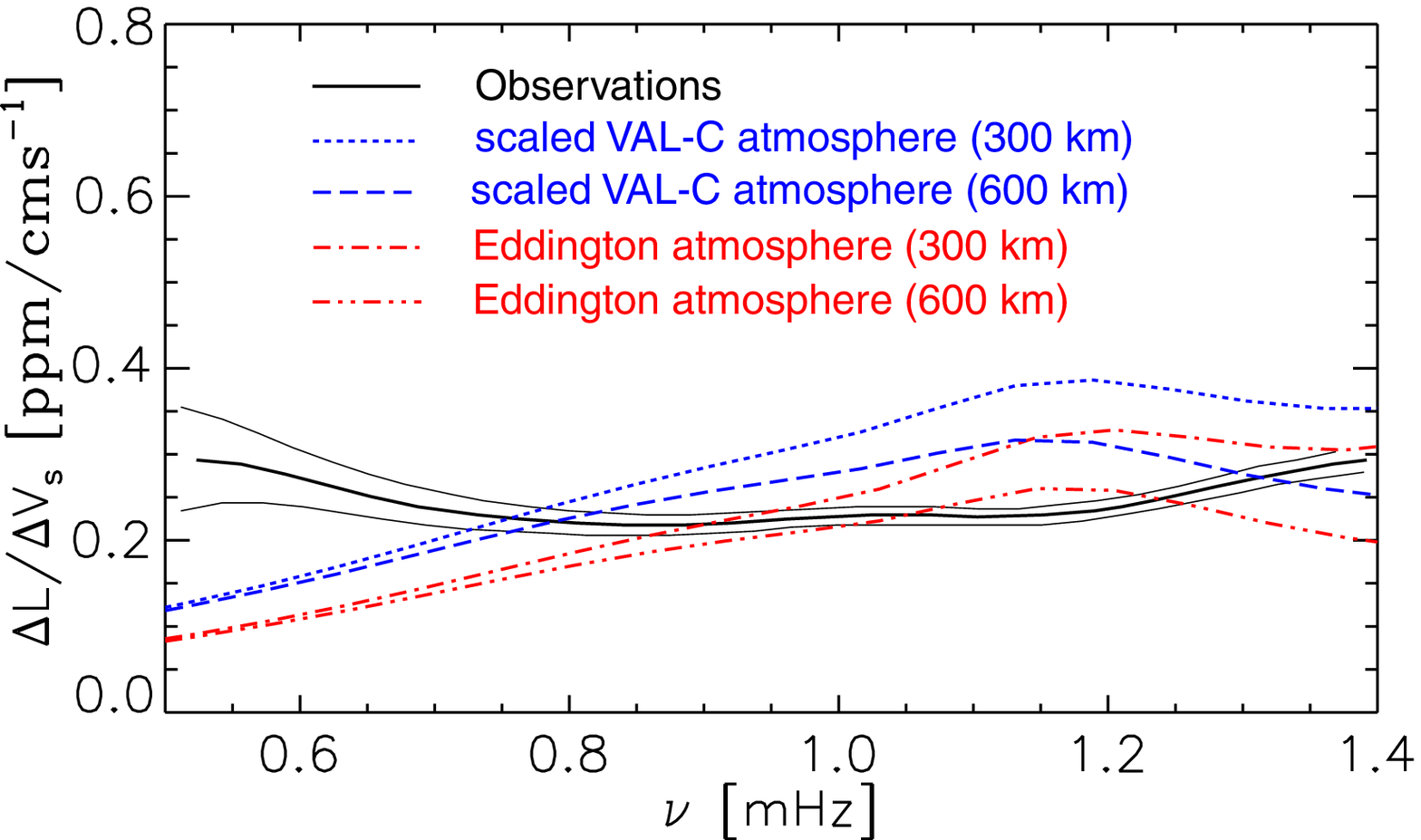}
\caption{Comparison of computed amplitude ratios of stochastically 
excited modes with observations. Left: results are shown for a 
solar model at different atmospheric heights (curves) and for the Sun (symbols).
The thick solid curve is a running mean average of the observed solar data
(adapted from Houdek et al. 1999). Right: results are shown for a model
of Procyon A at different atmospheric heights ($h$=300, 600 km) and for two different
atmospheres: a scaled VAL-C (Vernazza\et 1981) and an Eddington atmosphere. 
The solid curves (thin curves indicate 1$\,\sigma$ uncertainties) are 
the observations (Huber et al. 2011a).
} 
\label{f:4}
\end{figure}


\subsection{Pulsation eigenfunctions}
A very promising test of the pulsation theory, and in particular of the pulsation eigenfunctions, 
independent of an excitation model, is provided by comparing estimated intensity-velocity 
amplitude ratios, $\Delta L/\Delta V_{\rm s}$, with observations 
(e.g. Houdek et al. 1995, 1999; Jimenez et al. 1999; Jimenez 2002; Houdek 2009). The 
estimated amplitude ratios are obtained from taking the ratios of the complex pulsation 
eigenfunctions provided by the nonadiabatic stability computations, which must 
include the coupling of the turbulent velocity field with the oscillations.
A comparison of solar observations (Schrijver\et 1991) with model predictions are 
displayed in the left panel of Figure\,\ref{f:4}, where the model results are depicted for 
velocity amplitudes computed at different atmospheric heights $h$. The square root of the
 mode kinetic energy per unit increment of radius $r$, which is proportional 
to $r\rho^{1/2}\xi$, 
increases rather slowly with height; the density $\rho$, however, decreases very 
rapidly and consequently the displacement eigenfunction $\bm{\xi}$ increases and
the amplitude ratio decreases with height $h$.

In the right panel of Figure\,\ref{f:4} model results for the F5 star Procyon A are compared 
with observations (Guenther et al. 2008) from the MOST spacecraft and contemporaneous 
radial velocity measurements by Arentoft et al. (2008). Theoretical results are shown 
for two stellar atmospheres at two different atmospheric 
heights. For both stellar atmospheres the agreement 
with the observations is less satisfactory than in the solar case, indicating that 
we do not represent correctly the shape of the pulsation eigenfunctions in the outer 
stellar layers. Moreover, the modelled amplitude ratios significantly depend on the 
adopted atmospheric model. Consequently there is need for adopting more realistically 
computed atmospheres in the equilibrium models, particularly for stars with much 
higher surface temperatures than the Sun.

\begin{figure}[!ht]
\plotone{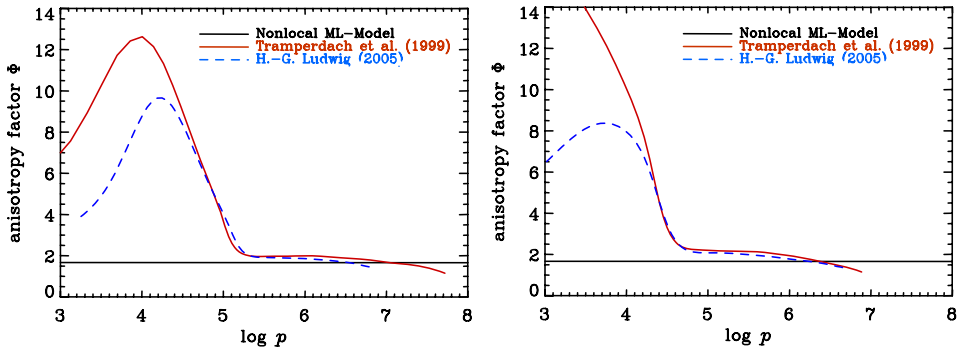}
\caption{
Anisotropy $\Phi$ of the turbulent velocity field as a function of the depth 
variable pressure ($p$) in the convectively unstable surface layers for models 
of the Sun (left panel) and Procyon A (right panel). Results are displayed for 
two different numerical simulations (red solid curve: R. Trampedach 1999, 
personal communication;
blue dashed curve: H.-G. Ludwig 2005, personal communication) and for our current version 
of a semi-analytical convection model of Gough (1977a,b; black horizontal line).
Figure adopted from Houdek (2011).
}
\label{f:5}
\end{figure}

Additionally to the effect of the atmospheric structure, the nonadiabatic 
eigenfunctions are also crucially modified by the dynamics of the turbulent 
velocity field in the convectively unstable surface layers. In particular the 
poorly modelled anisotropy of the turbulent velocity 
field ${\bm u}=(u,v,w)$ in the near-surface layers crucially affects the magnitude and 
frequency-dependence of the amplitude ratios (Houdek 2011). In our current time-dependent 
convection model (Gough 1977a,b) the anisotropy
${\Phi}:= {\bm u}\!\cdot\!{\bm u}/w^2$ is 
parametrized by a constant value of order unity. Numerical simulations, however, indicate 
that $\Phi$ varies rapidly with height in the outer stellar layers, as illustrated in 
Figure~\ref{f:5} for models of the Sun (left panel) and Procyon A (right panel). It therefore 
behoves us, 
with the help of numerical simulations, to develop a model for the velocity anisotropy 
$\Phi$ for 
a more realistic description of the shapes of the convective cells as a function of 
stellar radius
and consequently its effect on the pulsation eigenfunctions and frequencies.


\subsection{Linear damping rates}
Another uncertainty in determining amplitudes of solar-like pulsations in 
hotter stars are the modelling of the mode lifetimes (Houdek 2006, 
Chaplin\et 2009, Baudin et al. 2011).
Although an acceptable agreement between observed heights of solar low-degree modes 
in the Fourier spectrum and theoretical expectations has been achieved with the help 
of a time-dependent convection model (Houdek\et  1999, Houdek \& Gough 2002) the 
largest error in the theory lies in the calculation of the damping rates of low-order 
modes (Houdek\et 2001, Chaplin\et 2005, 2009). We located
our principal problem in the stability 
analyses to be in the convection zone, below the upper boundary layer. An even larger 
discrepancy between observationally and theoretically inferred acoustic spectral 
linewidths was reported for solar-like stars that are hotter than the Sun (e.g., Houdek 2006, 2009). 
For example, with the help of hydrodynamical simulations (Stein\et 2004), it was 
found that the theory underestimates the damping rates of the most prominent modes by a 
factor of two, and possibly more, for the F5 solar-like star Procyon (Houdek 2006). 
This failure in the theory has 
recently been confirmed by using average linewidth measurements of solar-like oscillations 
in several hundreds of stars observed by Kepler (Chaplin 2010, personal communication, see
also Appourchaux\et 2012). 

Current theories (Chaplin\et 2009) 
predict the average mode linewidth to scale with the fourth power of the stars effective 
temperature, whereas preliminary Kepler data suggest a steeper dependence, i.e. larger than 
four (see also Baudin\et 2011 using CoRoT, and Appourchaux\et 2012 using Kepler data). 
As in the case of Procyon, this indicates 
a failure of our current theory to predict mode linewidths in stars hotter than the Sun, most 
likely because of a physical mechanism still missing in our current theory. One such crucial 
mechanism is incoherent scattering at the inhomogeneous upper boundary layer (Murray 1993; 
Goldreich \& Murray 1994), which becomes increasingly more important for stars with higher 
masses and effective temperatures.


\begin{figure}[!ht]
\plotfiddle{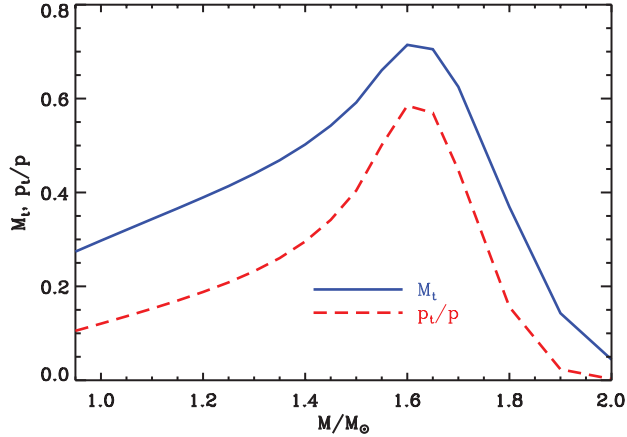}{5.5cm}{0}{100}{100}{-100}{00}
\caption{
Maximum values of the turbulent Mach number $M_{\rm t}:=w/c$ and
turbulent pressure $p_{\rm t}:=\langle\rho ww\rangle$ (angular brackets
denote an ensemble average) versus stellar mass for models along the 
ZAMS (adapted from Houdek \& Gough 1998).} 
\label{f:6}
\end{figure}


Murray (1993) and Goldreich \& Murray (1994) were one of the first to have stressed the 
importance of the contribution of incoherent scattering ($\eta_{\rm scatt}$) to the 
mode linewidths, 
and that it may dominate over other agents contributing to mode damping. They derived 
in the geometrical optics limit, at the top of the convective envelope, the scattering
contribution 
\begin{equation}
\eta_{\rm scatt}\sim\frac{\omega_{\rm r} M^2_{\rm t}}{\pi(n+1)} 
\end{equation}
for a radial mode with 
order $n$ and frequency $\omega_{\rm r}$, which is proportional to the squared 
turbulent Mach number $M_{\rm t}$.

It was recognised by Houdek \& Gough (1998) that the turbulent Mach number 
increases rapidly with stellar mass and surface temperature (see Figure~\ref{f:6}). 
Consequently the scattering 
contribution to the mode linewidths may therefore dominate over the other contributions 
for stars with high surface temperatures, or more precisely, for stars with turbulent 
Mach numbers considerably larger than in the Sun. Hot stars have relatively shallow 
surface convection zones but very vigorous turbulent velocity fields 
(e.g. Houdek \& Gough 1998). 

Goldreich \& Murray (1994) adopted in their scattering model a time-independent, 
mixing-length-like, 
convection model and concluded that the mode energy is most effectively scattered 
into other modes of similar frequency but higher spherical degree, with the surface 
gravity mode (f mode) being the ultimate recipient of the scattered energy. 
A time-dependent treatment of the 
turbulent velocity field will increase the frequency range of the modes over which the 
pulsation energy can be scattered (Gough 1977a). An immediate question to be asked here 
is what the frequency 
distribution of this process will be, because the distribution determines not only the mode 
linewidths but also the shape of the spectral peaks in the oscillation power spectrum 
(e.g., Rast \& Gough 1995; Rast 1999). Moreover, if the frequency behaviour of the damping 
process is similar to that of the stochastic excitation process, which is very likely the 
case (e.g. Houdek\et 1999), we would be in the position to put further constraints on 
the frequency factor of the turbulent velocity spectrum in any stochastic excitation model 
(e.g. Houdek 2010).

\acknowledgements I am thankful to Reza Samadi for providing some of the numbers listed
in Table~1 and to the KITP staff of UCSB for their warm hospitality
during the research programme ``Asteroseismology in the Space Age''.
This work was supported by the Austrian Science Fund (FWF), Project P21205-N16,
and in part by the National Science Foundation of the United States
under Grant No.\ NSF PHY05--55164.


\end{document}